\documentclass[12pt,a4paper]{article}
\usepackage{amsmath}
\usepackage{amssymb}
\usepackage{cite}
\usepackage{hyperref}

\title{ Do   the uncertainty relations  really have crucial significances for physics ?}
\author{Spiridon  DUMITRU \footnote{(ret.from)Department of Physics , \textit{"`Transilvania"'} University, B-dul Eroilor 29, 500036 Brasov,  Romania, e-mail: s.dumitru42@yahoo.com  }}

\date{\today}

\begin{document}

\maketitle

\begin{abstract}
It is proved  the falsity of  idea that the  Uncertainty Relations (UR) have  crucial significances for physics. Additionally  one argues for the necesity of an  UR-disconnected quantum philosophy.
\end{abstract}
\section{Introduction}
The Uncertainty Relations (UR) enjoy  a considerable popularity, due in a large measure    to the so called Conventional (Copenhagen)  Interpretation of UR (CIUR). The mentioned popularity is  frequently associated  with the idea (which persist so far) that UR have   
 crucial significances for physics (for a list of relevant references see  \cite{1,2,3}). The   itemization of the alluded idea  can be done through the following more known Assertions 
( $\textbf{\textit{A}}$ ): 

$\bullet$ $\textbf{\textit{A}}_1$ :  In an experimental reading the UR are  crucial  symbols  for measurement characteristics regarding Quantum Mechanics (QM)  in contrast  with non-quantum Classical  Physics  (CP). The pointed characteristics view  two aspects : (i) the so called 'observer effect' (i.e. the perturbative influence of 'observation'/measuring devices  on the investigated system), and (ii) the measurement errors (uncertainties). Both of the alluded   aspects are presumed to be  absolutely notable and unavoidable in QM contexts respectively entirely negligible and avoidable in CP situations. $\rule{0.15cm}{0.15cm}$ 

$\bullet$ $\textbf{\textit{A}}_2$ : From a theoretical viewpoint UR are essential distinction
elements between the theoretical frameworks of QM and  CP. This in sense of the supposition that mathematically UR appear only in QM pictures and have not analogues  in the  CP representations. $\rule{0.15cm}{0.15cm}$ 

$\bullet$ $\textbf{\textit{A}}_3$ : In both experimental and theoretical  acceptions the UR are in an indissoluble
connection with the   description of uncertainties  (errors) specific for Quantum Measurements (QMS). $\rule{0.15cm}{0.15cm}$ 

$\bullet$ $\textbf{\textit{A}}_4$ : As an esential piece of UR, the Planck's constant $\hbar$, is appreciated  to be exclusively
 a symbol of quanticity (i.e. a signature of QM comparatively with CP ),  without any kind of analogue in CP. $\rule{0.15cm}{0.15cm}$ 

$\bullet$ $\textbf{\textit{A}}_5$ : UR  entail \cite{4} the existence of some  'impossibility' (or 'limitative') principles  in foundational physics . $\rule{0.15cm}{0.15cm}$ 

$\bullet$ $\textbf{\textit{A}}_6$ : UR are regarded \cite{5} as  expression of \textit{"`the most important principle of the twentieth century physics"'}. $\rule{0.15cm}{0.15cm}$ 

 To a certain extent the verity of the idea itemized by assertions $\textbf{\textit{A}}_1 - \textbf{\textit{A}}_6$ depends on the entire  truth of CIUR. That is why in the next section we present briefly the CIUR untruths which  trouble the mentioned verity. Subsequently, in Section 3, we point out a lot of Observations ($\textbf{\textit{O}}$) which invalidate completely and irrefutably  the items    $\textbf{\textit{A}}_1 - \textbf{\textit{A}}_6$. The respective invalidation suggests a substitution of UR-subordinate quantum  philosophy with an UR-disconnected conception. Such a suggestion is consolidated by some additional Comments ($\textbf{\textit{C}}$) given in Section 4. So, in  Section 5, we can conclude our  considerations with: (i) a  definitely negative answer to the inquired idea, respecively (ii) a pleading for a new quantum philosophy. Such conclusions argue for the Dirac's intuitional guess  about the non-survival of  UR in the physics of future.

\section{Shortly on the CIUR untruths}
In its essence the  CIUR doctrine was  established  and disseminated  by the founders and subsequent partisans of Copenhagen School in QM. The story  started from the wish  to give out  an unique and  generic interpretation for the thought-experimental (\textit{te}) formula
\begin{equation}\label{eq:1}
\Delta _{te} A \cdot \Delta _{te} B \ge \hbar 
\end{equation}
( $A$ and $B$ being  conjugated observables) respectively for the QM theoretical formula
\begin{equation}\label{eq:2}
\Delta _\psi  A \cdot \Delta _\psi  B \ge \frac{1}{2}\left| {\left\langle 
{\left[ {\hat A,\hat B} \right]} \right\rangle _\psi  } \right|
\end{equation}
(where the notations are the usual ones from usual QM - see also \cite{3}). Both the above two kind of formulas are known as UR.

The alluded doctrine remains  a widely adopted conception  which, in various manners,  dominates  to this day  the questions regarding the foundation and interpretation of QM.   However, as a rule, a minute  survey of the  truths-versus-untruths
regarding its substance    was (and still is) underestimated in the main stream of  publications (see the literature mentioned in \cite{1,2}). This in spite of the early known  opinions like \cite{6}  : \textit{"`the idea that there are defects in the foundations of orthodox quantum theory is unquestionable present in the conscience of many physicists"'}.

A survey of the mentioned kind  was   approached by us in the report \cite{3} as well as in
its precursor  papers \cite{7,8,9,10,11,12,13,14,15} and preprints \cite{16}. Our approaches, summarized in \cite{3},
disclose the fact  that each of all   basic elements (presumptions) of CIUR are troubled by a number of insurmountable shortcomings (unthruths).   For that reason
 we believe that  CIUR must be wholly abandoned  as a wrong construction which, in its substance, has no noticeable
 value for physics. The disclosures from \cite{3}  
 were  carried out   by an entire class of well argued remarks ($\textbf{\textit{R}}$). From the mentioned  class  we compile here only the following ones:

$\bullet$ $\textbf{\textit{R}}_a$ : Formula \eqref{eq:1} is mere  provisional fiction without any  durable physical significance.  This because  it has only  a transitory/temporary character,  founded on old  resolution criteria from optics  (introduced by Abe and Rayleigh). But the respective criteria were surpassed by the so called super-resolution techniques worked out in modern experimental physics. Then, instead of CIUR formula \eqref{eq:1}, it is possible to imagine some 'improved relations' (founded on some super-resolution thought-experiments) able to invalidate in its  very essence the respective formula. $\rule{0.15cm}{0.15cm}$ 

$\bullet$ $\textbf{\textit{R}}_b$ : From a theoretical perspective the formula \eqref{eq:2} is only   a minor and deficient piece,
 resulting  from the genuine  Cauchy-Schwarz relation 
\begin{equation}\label{eq:3}
\Delta _\psi  A \cdot \Delta _\psi  B \ge \left| {\left( {\delta _\psi  \hat A\,\psi ,\delta _\psi  \hat B\,\psi } \right)} \right|
\end{equation}
written in terms of usual QM notations (see\cite{3}).\\ 
As regards their physical significance the formulas \eqref{eq:2} and\eqref{eq:3} are  nothing but simple (second order) fluctuations relations  from the same family with the similar ones \cite{3,7,8,9,12,15} from the
 statistical  CP . $\rule{0.15cm}{0.15cm}$ 

$\bullet$ $\textbf{\textit{R}}_c$ : In a true approach the formulas \eqref{eq:1} and \eqref{eq:2} as well as their  'improvised adjustments' have no connection with the description of QMS. $\rule{0.15cm}{0.15cm}$ 
    
$\bullet$ $\textbf{\textit{R}}_d$ : The Planck's constant $\hbar$ besides its well-known quanticity significance is endowed also \cite{3,12}  with the quality of generic indicator for quantum randomness (stochasticity) - i.e. for the   random characteristics of  QM  observables. Through such a quality $\hbar$ has \cite{3,12} an authentic  analogue in statistical CP. The respective analogue is  the Boltzmann's constant $k_B$ which is an authentic generic indicator for thermal randomness. Note that, physically, the randomness of an observable is manifested  through its fluctuations \cite{3,7,8,9,12,15}. $\rule{0.15cm}{0.15cm}$ 

$\bullet$ $\textbf{\textit{R}}_e$ :
The formula \eqref{eq:2} is not applicable
for  the pair of  (conjugated) observables $t - E $ (time - energy ). In other words \cite{3}
  a particularization  of  \eqref{eq:2} in the form
\begin{equation}\label{eq:4}
\Delta _\psi  t \cdot \Delta _\psi  E \ge \frac{\hbar }{2}
\end{equation}
gives in fact a wrong relation. This because in usual QM the time \textit{t} is a
deterministic variable but not a random one. Consequently for any QM situation
 one finds the expressions $\Delta_{\psi}t \equiv 0$ respectively $\Delta_{\psi}E = a\; finite\; quantity$. \\
Note that in a correct mathematical-theoretical approach for the $t -E$ case it  is valid only the  Cauchy Schwarz formula  \eqref{eq:3}, which  degenerate into trivial relation $0 = 0$. 
 $\rule{0.15cm}{0.15cm}$ 

Starting from the above  remarks  $\textbf{\textit{R}}_a - \textbf{\textit{R}}_e$ in the next section we add an entire
 group of Observations ($\textbf{\textit{O}}$ ) able to give 
 a just estimation of correctness
 regarding the assertions $\textbf{\textit{A}}_1 - \textbf{\textit{A}}_6$.
\section{The falsity of assertions $\textbf{\textit{A}}_1 - \textbf{\textit{A}}_6$ }
The  above announced  estimation   can be obtained only if the mentioned remarks are supplemented with some other notable elements. By such a supplementation one obtains   a  panoramic  view which can be reported through  the whole group of the following   Observations ($\textbf{\textit{O}}$) :

$\bullet$ $\textbf{\textit{O}}_1$ : The remark  $\textbf{\textit{R}}_a$, noted in previous section, shows irrefutably the falsity of the assertion $\textbf{\textit{A}}_1$. The same falsity is argued by the fact that the referred 'observer effect' and corresponding measuring uncertainties can be noticeable not only in QMS  but also  in some   CP measurements (e.g. \cite{17} in electronics or in thermodynamics)  $\rule{0.15cm}{0.15cm}$   

$\bullet$ $\textbf{\textit{O}}_2$ : On the other hand the remark $\textbf{\textit{R}}_b$ points out the evident  untruth of the assertion $\textbf{\textit{A}}_2$. $\rule{0.15cm}{0.15cm}$ 

$\bullet$ $\textbf{\textit{O}}_3$ : Furthermore the triplet of remarks $\textbf{\textit{R}}_a - \textbf{\textit{R}}_c$ infringes the essence of the assertion $\textbf{\textit{A}}_3$. $\rule{0.15cm}{0.15cm}$ 

$\bullet$ $\textbf{\textit{O}}_4$ : The exclusiveness feature of Planck's constant $\hbar$, asserted  by  $\textbf{\textit{A}}_4$, is evidently  contradicted by the remark $\textbf{\textit{R}}_d$. $\rule{0.15cm}{0.15cm}$

$\bullet$ $\textbf{\textit{O}}_5$ : Assertion  $\textbf{\textit{A}}_5$  was reinforced 
and disseminated recently \cite{4} thrugh the topic:
\begin{quote}
\textit{"`What role do 'impossibility' principles or other limits (e.g., sub-lightspeed signaling, \textbf{Heisenberg uncertainty}, cosmic censorship, the second law of thermodynamics, the holographic principle, \textbf{computational limits}, etc.) play in foundational physics and cosmology?"'}. 
\end{quote}
Affiliated oneself with the quoted topic the assertion  $\textbf{\textit{A}}_5$  implies two readings: (i) one which hints at
Measuring Limits (ML), respectively (ii) another associated with  the so called 'Computational Limits'(CL). $\rule{0.15cm}{0.15cm}$

$\bullet$ $\textbf{\textit{O}}_6$ : In the  reading connected with ML the assertion $\textbf{\textit{A}}_5$ presumes that  the  QMS accuracies can not surpass 'Heisenberg uncertainties' \eqref{eq:1} and \eqref{eq:2}. Such a presumption is perpetuated until these days through sentences like : \textit{"`The uncertainty principle of quantum mechanics  places a fundamental limitation
on what we can know"'} \cite{18}.\\ 
Now is easy to see that the above noted remarks $\textbf{\textit{R}}_a$ and $\textbf{\textit{R}}_c$ reveal beyond doubt
 the weakness of such a presumption. Of course that, as a rule, for various branches of physics (even of CP nature such are  \cite{17} those from electronics or thermodynamics), the existence  of some  specific ML is a  reality. The respective existence is subordinate to certain genuine elements such  are the accuracy of experimental devices and the competence of the theoretical approaches. But note that as it results from the alluded remarks the formulas  \eqref{eq:1} and \eqref{eq:2} have nothing to do with  the evaluation or description of the  ML (non-performances or uncertainties) regarding QMS . $\rule{0.15cm}{0.15cm}$

$\bullet$ $\textbf{\textit{O}}_7$ :  The  reading   which associate the UR with CL sems to refer mainly to the  Bremermann's limit (i.e. to the maximum computational speed of a  self-contained system in the  universe) \cite{19,20}. But it is easy to see  from \cite{19,20} that the aludded association is builded in fact  on the wrong relations \eqref{eq:1}and \eqref{eq:4} written for the observables pair \textit{t - E}.  Consequently
such an association has not any real value for appreciation of UR significance as CL. Add here the remark that, nevertheless, the search \cite{20} for finding the ultimate physical limits of computations  remains  a subject worthy to be investigated. This because, certainly, that what is ultimately permissible in practical computational progresses
depends on what are the ultimate possibilities  of real physical artifacts (experiences). However, from our viewpoint, appraisals of the alluded possibilities  do not require any appeal to the  the relations \eqref{eq:1}, \eqref{eq:2} or \eqref{eq:4}. $\rule{0.15cm}{0.15cm}$

$\bullet$ $\textbf{\textit{O}}_{8}$ : For a true  judgment  regarding the validity of assertion $\textbf{\textit{A}}_6$   can be taken into account    the following aspects:
\begin{quote}
(i)  In its essence  $\textbf{\textit{A}}_6$ prove oneself to be nothing but an unjustifiable distortion of the real  truths. Such a proof  results directly  from the  above remarks $\textbf{\textit{R}}_a - \textbf{\textit{R}}_c$. According to the alluded remarks in reality the UR \eqref{eq:1} and \eqref{eq:2} are  mere provisional fictions  respectively minor (and restricted) QM relations. So it results that, in the main,
 UR are insignificant things comparatively with the true important principles of the 20th century physics (such are  the ones regarding  \textit{Noether's theorem}, \textit{mass-energy equivalence}, \textit{partricle-wave duality} or \textit{nuclear fission} ).

(ii) It is wrongly to promote the assertion $\textbf{\textit{A}}_6$ based  on  the existent publishing situation where, in the mainstream
of QM text-books, the UR \eqref{eq:1} and/or \eqref{eq:2} are amalgamated with the basic quantum concepts. The wrongness is revealed by the fact that the alluded situation  was created through an unjustified perpetuance of the writing style  done by the CIUR partisans.

(iii) The  assertion $\textbf{\textit{A}}_6$ must be not  confused with
  the history confirmed remark \cite{21} :
\textit{UR "`are probably the most controverted formulae in
the whole of the theoretical physics"'}. With more justice the respective remark  has to be regarded as  accentuating the weakness of  concerned assertion. 
\end {quote}
Together the three above noted aspects give enough reasons for an incontestable
incrimination of the assertion under dicussion.   $\rule{0.15cm}{0.15cm}$ 

The here detailed observations $\textbf{\textit{O}}_1 - \textbf{\textit{O}}_8$ assure   sufficient solid arguments in order to prove the indubitable incorrectness for each of  the assertions $\textbf{\textit{A}}_1 - \textbf{\textit{A}}_6$  and, consequently, the falsity of the idea that UR really have crucial
significances for physics. But the alluded  
 proof conflicts with the UR-subordinate 
 quantum philosophy in  which the interpretational questions of QM and debates about QMS description are  indissolubly associated with the formulas \eqref{eq:1} and/or \eqref{eq:2}. The true (and deep) nature of the  respective conflict  suggests directly the necesity of  improvements  by  substituting the alluded philosophy  with another UR-disconnected conception.
 
Of course that the before-mentioned substitution necessitates further well argued reconsiderations, able to gain the support of mainstream  scientific communities and publications.  Note that, in one way or other, elements of the UR-subordinate  philosophy are present  in almost all current QM interpretations \cite{22,23}. We think that among the  possible multitude  of elements belonging  to the  alluded reconsiderations  can be included  the  additional group of comments from the next section.

\section{Some additional comments}
The Comments ($\textbf{\textit{C}}$) from the foregoing  announced group,  able to suggest also improvements  in quantum philosophy, are the following ones:

$\bullet$ $\textbf{\textit{C}}_1$ : Firstly we   
note  that  the substance of above presented remarks  $\textbf{\textit{R}}_a - \textbf{\textit{R}}_b$ respectively observations   $\textbf{\textit{O}}_1 - \textbf{\textit{O}}_3$ can be  fortified  by means of the following three our views:
\begin {quote}
 (i) In its bare and lucrative framework, the usual QM offers solely theoretical models for  own  characteristics of the investigated systems (microparticles of atomic size).
 
(ii) In the alluded framework  QM has no connection with a natural depiction of QMS.

(iii) The  description of QMS   is an autonomous subject, investigable   in addition to  the  bare theoretical structure of usual QM.
 \end{quote}
  We think that, to a certain extent, our above  views find  some support in the Bell's remark \cite{24} : \textit{"`the word (\textbf{'measurement}') has had such a damaging efect on the discussions that . . . it should be banned altogether in quantum mechanics "'}. (It happened that, in a  letter \cite{25}, J.S.Bell comunicated us  early the essence of the  alluded remark together with a short his personal  agreement with our incipient opinions about UR and QMS). $\rule{0.15cm}{0.15cm}$
  
$\bullet$ $\textbf{\textit{C}}_2$ : In its substance the view (i) from $\textbf{\textit{C}}_1$ regards the  bare QM as being  nothing but an abstract (mathematical) modeling of the properties  specific to the atomic-size sytems (microparticles). For a given system the main elements of the alluded modeling are the wave functions 
$\psi_\alpha$, respectively the quantum operators $\hat{A}_j$. On the one hand   $\psi_\alpha$ describes the  probabilistic situation of the system in    $\alpha$ state. Mathematically $\psi_\alpha$ is nothing but the solution of the corresponding Schrodinger equation. On the other hand each of the operators $\hat{A}_j$ ($j = 1,2,...,n$) is a generalised radom variable associated to a specific observable $A_j$ (e.g. coordinate, momentum, angular momentum or energy) of the system. Then in a probabilistic sense the global characterization of the observables $A_j$ is given by the expected parameters: \\
 (i) the mean values $\left\langle {A_j } \right\rangle _\psi   = \left( {\psi ,\hat A_j \psi } \right)$ wherre $\psi  \equiv \psi _\alpha $ while $\left( {f,g} \right)$ denotes the scalar product of functions $f$ and $g$, \\
 (ii) the ($r+s$)-order correlations ${\rm K}_\psi  \left( {i,j;r,s} \right) = \left( {\left( {\delta _\psi  \hat A_i } \right)^r \psi ,\left( {\delta _\psi  \hat A_j } \right)^s \psi } \right)$,  with $\delta _\psi  \hat A_j  = \hat A_j  - \left\langle {A_j } \right\rangle _\psi$  and $r + s \ge 2$.\\
So the definitions of parameters $\left\langle {A_a } \right\rangle _\psi$ and ${\rm K}_\psi  \left( {i,j;r,s} \right)$ appeal to the usual notations from known  QM texts (see \cite{3,26,27}). $\rule{0.15cm}{0.15cm}$

$\bullet$ $\textbf{\textit{C}}_3$ :
The before mentioned QM entities are completely similar with the known things from statistical CP (such are  the phenomenological theory of fluctuations  \cite{28,29} respectively the classical statistical mechanics \cite{30,31}). So the wave functions $\psi_\alpha$ correspond to the   probability distributions
$w_\alpha$ while the operators $\hat{A}_j$ are alike the macroscopic random observables $\mathbb{A}_j$. Moreover the QM  probabilistc expected parameters $\left\langle {A_a } \right\rangle _\psi$ and ${\rm K}_\psi  \left( {i,j;r,s} \right)$ are entirely analogous with the mean values respectively the second and higher order fluctuations correlations regarding the  macroscopic observables $\mathbb{A}_j$
 \cite{3,7,8,9,10,12,15,28,29,30,31,32}
$\rule{0.15cm}{0.15cm}$

$\bullet$ $\textbf{\textit{C}}_4$ : It is interesting to complete the  above comment with the following annotations. Undoubtedly that, mathematically,  the QM  observables have innate characteristics of random variables. But similar characteristics one finds also in the case  of statistical CP observables. Then it is surprisingly that     
the two kinds of random observables (from QM and CP)  in their connection  with the problem of measurements are approached  differently by the same  authors \cite {26,30}  or teams \cite{27,31}. Namely the alluded problem is  totally neglected in the case of CP observables \cite{30,31}, respectively it is regarded  as a capital question for QM observables \cite{26,27}. Note that the mentioned differentiation is not justified  \cite{26,27,30,31} by any physical argument. We think that, as regard te description of their measurements, the two kinds of random observables must be approached in similar manners.\\
In the context of above annotations it is  interesting to mention the following very recent statement \cite{33} : \textit{"`To our best current knowledge the measurement process in quantum mechanics is non-deterministic"'}.  The inner nature of the mentioned statement strengthens our appreciation \cite{3} that   a measurement of  a (random) quantum observable must be understood   not as
a single trial (which give a unique value) but as a statistical selection/sampling (which yields a spectrum of values). Certainly that in such an understanding the concept  of 'wave function collapse' \cite {34,35} becomes an obsolete thing.
$\rule{0.15cm}{0.15cm}$ 

$\bullet$ $\textbf{\textit{C}}_5$ : 
A credible tentative in  approaching similarly the description  of measurements regarding random observables from both QM and CP was promoted by us in \cite{3,36}.  Our  approach was done according the views (ii) and (iii) noted in the above comment $\textbf{\textit{C}}_1$. Mainly the respective approach aims to obtain a well argued (and consequently credible) description of QMS.  So,   in papers \cite{3,36}, a QMS was depicted as a distortion  of the information about the measured system.  For a given system the respective distortion  is described (modeled) as  a process which change linearly the   probability density and current (given in terms of wave function) but preserve the mathematical expressions of QM operators regarded as generalised random variables. Note that an  analogous  description of measurements concerning the  random observables from CP was done by us formerly in \cite{37}.  
$\rule{0.15cm}{0.15cm}$

$\bullet$ $\textbf{\textit{C}}_{6}$ : Other open   question  of quantum philosophy regards the    deterministic subjacency  of QM randomness. The  question,  of great interest\cite{38}, aims  to clarify  if the respective randomness has an irreducible   nature or otherwise it derives from the existence of  some subjacent  hidden variables of deterministic essence. Then it appears as a notable aspect the fact that, in so reputable  report \cite{38}  about the alluded question, the possible involvement of UR \eqref{eq:1} and/or \eqref{eq:2} is completely omited. Such a remarkable omission show clearly that the UR \eqref{eq:1} and/or \eqref{eq:2} do not present any interest for one of the most thought-provoking subject regarding quantum philosophy.
$\rule{0.15cm}{0.15cm}$

$\bullet$ $\textbf{\textit{C}}_{7}$ :
Here is the place to refer comparatively  to the deterministic subjacency regarding CP kind of randomness. The respective kind is associated (both theoretically and experimentally)  with a class of subjacent deterministic variables,  specific to the molecular and atomic motions \cite{28,29,30,31}. The important feature of the alluded  CP subjacency is  the fact that it does not annul at all the corresponding randomness. Namely the respective deterministic subjacency do not revoke at all the   random  entities such are the probability distributions
$w_\alpha$ and macroscopic observables $\mathbb{A}_j$ , mentioned above in $\textbf{\textit{C}}_{3}$.
 The respective entities keep the essence of the 
  CP randomness  revealed physically through the corresponding global  fluctuations of macroscopic observables .\\
We think that the  noted  classical feature must  be taken as a reference element in  managing the discussions regarding the deterministic subjacency of  QM (i.e. the question of hidden variables - versus - QM randomness) and,  generally speaking, the renovation   of quantum philosophy. More exactly it is of direct interest to see if the existence of hidden variables  removes or keeps the QM randomness incorporated within the wave functions $\psi_\alpha$ and operators $\hat{A}_j$. We dare to believe that the alluded QM randomness will persist, even if the existence of some subjacent hidden variables would be evidenced (first of all experimentally).  
$\rule{0.15cm}{0.15cm}$

$\bullet$ $\textbf{\textit{C}}_{8}$ : Now some other words about the question  of 'impossibility' principles  in foundational physics, discussed   above in observations $\textbf{\textit{O}}_{5}- \textbf{\textit{O}}_{7}$. The respective principles were mentioned in connection with questions like : ' What is Ultimately Possible in Physics?' (see \cite{4}). To a deeper analysis the alluded connection calls attention to 'the frontier of knowledge'. In scrutinizing the respective frontier  it was acknowledged recently \cite{33} that: 
\textit{"`Despite long efforts, no \textbf{progress }has been made}...for ...\textit{the understanding of
quantum mechanics, in particular its measurement process and interpretation"'}. What is most important in our opinion is the fact that, in reality, for the sought \textit{\textbf{"`progress"'}} the UR \eqref{eq:1} and \eqref{eq:2} are  of no interest or utility.  $\rule{0.15cm}{0.15cm}$

By ending this section it is easy to  see that the here added comments
$\textbf{\textit{C}}_{1} - \textbf{\textit{C}}_{8}$ give supports to the before suggested  proposal for a UR-disconnected quantum philosophy. 

\section{ Conclusions}
A  survey, in Section 3, of the observations  $\textbf{\textit{O}}_1 -  \textbf{\textit{O}}_8$   discloses that \textbf{\textit{in fact the UR  \eqref{eq:1} and \eqref{eq:2} have not any crucial significance for physics}}. Additionally, in Section 4,
an examination of the  comments
$\textbf{\textit{C}}_1 -  \textbf{\textit{C}}_8$
provides supporting elements   for a UR-disconnected quantum philosophy. 

So  we give forth a class of solid arguments which come to advocate and consolidate the Dirac's intuitional  
guess \cite{39} : \textit{\textbf{"`uncertainty relations in their present form will not survive in the physics of future"'}}. 

\end{document}